\documentclass[assymb,11pt]{article}
\usepackage{amssymb}
\newcommand{\be}{\begin{equation}}
\newcommand{\ee}{\end{equation}}
\newcommand{\ber}{\begin{eqnarray}}
\newcommand{\ear}{\end{eqnarray}}
\newcommand{\ba}{\begin{array}}
\newcommand{\ea}{\end{array}}

\newcommand{\de}{\delta}

\newcommand{\fr}{\frac}

\newcommand{\Li}{\pounds}
\newcommand{\n}{\nonumber\\}

\newcommand{\p}{\partial}
\begin{document}
\title{Interacting with the Fifth Dimension.}
\author{
Mark D. Roberts,\\
     {University of Surrey},  GU2 7XH,\\
            {m.d.roberts@surrey.ac.uk},
{http://www.ph.surrey.ac.uk/$\sim$phx1mr}}
\maketitle
\begin{abstract}
Some new five dimensional minimal scalar-Einstein exact solutions are presented.
These new solutions are tested against various criteria used to measure interaction
with the fifth dimension.
\end{abstract}
{\tiny\tableofcontents}
\section{Introduction}
The use of the scalar-Einstein field equations and their solutions seems to have
gone through three,  perhaps related,  stages.
In the {\it first} stage \cite{fischer} solutions were sought which it was hoped would
represent an elementary particle such as a meson.
How the scalar field decay from such solutions might effect the Yukawa potential
has been discussed in \cite{mdr39}.
In the {\it second} stage it was noticed that most scalar-Einstein solutions do not
have event horizons.
It was shown that no event horizons happens under fairly general conditions
in the static case \cite{chase}.
In the non-static case there are imploding solutions which create curvature
singularities out of nothing \cite{mdrphd} \cite{mdr13},
these examples have no overall mass,
but there is a mass in the corresponding conformal-scalar solutions \cite{mdr23}.
In the {\it third} stage exact scalar-Einstein solutions where found to be
critical cases in the numerical study of stellar collapse \cite{choptuik}.

Contemporary attempts at quantum gravity and unification usually
involve more than the observable four dimensions \cite{PT} \cite{wuensch}.
It is possible that models in five dimensions might provide
testable cosmological models \cite{RS} \cite{maartens},
or be testable on the scale of the solar system\cite{WML},
or perhaps be testable microscopically.
To build a five dimensional model
in addition to an exact solution one needs to prescribe another piece of information:
how the four dimensional spacetime is embedded in the five dimensional space.
This is typically done by requiring that the four dimensional spacetime
is a four dimensional surface in a five dimensional space,
this can be achieved by choosing a normal vector field $n^a=\de^a_\chi$ to the surface.
Once five dimensional scalar-Einstein solutions have been found there turns
out to be many inequivalent ways of doing this and it is not immediate which is the best.
For simplicity here mainly solutions to the field equations $R_{ab}=2\phi_a\phi_b$
are discussed.  In particular these field equations and spherical symmetry require
$R_{\theta\theta}=0$ so that there is no self-interaction, such as mass,  for the
scalar field,  and also there is no cosmological constant present,  both of these
require $R_{\theta\theta}~\alpha~ g_{\theta\theta}$.
The cosmological constant is often taken to be related to a brane tension,
so that the examples here are for zero tension.

Most calculations were done using GRTensorII/Maple9 \cite{MPL}.
\section{The four dimensional solution.}
In single null coordinates the line element is
\be
ds^2_4=-(1+2\sigma)dv^2+2dvdr+r(r-2\sigma v)d\Sigma^2_2,~~~
d\Sigma^2_2=d\theta^2+\sin(\theta)^2d\phi^2.
\label{esnd4}
\ee
The scalar field takes the form
\be
\phi=\fr{1}{2}\ln\left(1-\fr{2\sigma v}{r}\right).
\label{sfsn4}
\ee
To transformation to double null coordinates use
\be
u\equiv(1+2\sigma)v-2r,
\label{dncoord}
\ee
so that the line element becomes
\be
ds^2_4=-dudv+Y^2d\Sigma^2_2,~~~
Y^2\equiv\fr{1}{4}((1+2\sigma)v-u)((1-2\sigma)v-u)
\label{ednd4}
\ee
where $Y$ is the luminosity distance.
In double null coordinates the scalar field takes the form
\be
\phi=\fr{1}{2}\ln\left(\fr{(1-2\sigma)v-u}{(1+2\sigma)v-u}\right),
\label{dn4s}
\ee
and the Ricci scalar is simply expressed
\be
R=\fr{2\sigma^2uv}{Y^4},
\label{rs4}
\ee
using the preferred vector field $v^a=\de^a_v$,
the magnetic part of the Weyl tensor vanishes,
and the electric part has one component
\be
E_{vv}=\fr{1}{3}R,
\ee
with $\Li_v E_{ab}=0$.
The product invariants can be expressed in terms of the Ricci scalar
\be
Rie^2=3R^2,~~~
Ricci^2=R^2,~~~
Weyl^2=\fr{4}{3}R^2,
\ee
as can the Carminati-McLenaghan \cite{CM} invariants
\ber
&R_1=3.2^{-4}R^2,~~~
R_2=-3.2^{-6}R^3,~~~
R_3=3.7.2^{-10}R^4,\\
&\Re(W_1)=3^{-1}.2^{-1}R^2,~~~
\Re(W_2)=3^{-2}.2^{-2}R^3,\n
&\Re(M_2)=M_3=3^{-1}.2^{-5}R^4,~~~
M_4=3^{-1}.2^{-8}R^5,~~~
\Re(M_5)=3^{-2}.2^{-6}R^5,\n
&\Im(W_1)=\Im(W_2)=\Re(M_1)=\Im(M_1)=\Im(M_2)=\Im(M_5)=\Re(M_6)=\Im(M_6)=0.\nonumber
\label{mc}
\ear
\section{The first five dimensional solution.}
In double null coordinates the first five dimensional generalization
of the four dimensional solution is
\ber
&ds^2_5=-\fr{1}{1-\alpha\beta}dudv+Y^2d\Sigma^2_2+d\chi^2,\\
&4Y^2\equiv((1+2\sigma)v-u+2\alpha\chi)((1-2\sigma)v-u+2\beta\chi),\n
&\phi=\fr{1}{2}\ln\left(\fr{(1-2\sigma)v-u+2\beta\chi}{(1+2\sigma)v-u+2\alpha\chi}\right),
\nonumber
\label{e5d94}
\ear
Defining
\be
\gamma\equiv(1-2\sigma)\alpha-(1+2\sigma)\beta
\label{gamma}
\ee
the Ricci scalar is given by
\be
R=
[16\sigma^2(1-\alpha\beta)uv+((\alpha-\beta)u-\gamma v)^2
+4(1-\alpha\beta)(2\sigma((\alpha-\beta)u+\gamma v)+(\alpha-\beta)\gamma\chi)\chi]/(8Y^4)
\label{r5e5d94}
\ee
The product invariants are again completely determined by the Ricci scalar,
however this time
\be
{Rie}^2=3R^2,~~~
{Ricci}^2=R^2,~~~
{Weyl}^2=\fr{11}{6}R^2,
\label{5dinvars}
\ee
so that the fraction multiplying the Weyl invariant has increased by $1/2$
from $4/3$ to $11/6$,  signifying that there is more gravitational field present.

Projecting onto a surface
\be
n_a=\de_a^\chi,
\label{proj}
\ee
which has vanishing acceleration and rotation,
and expansion, shear and extrinsic curvature given by
\ber
&\Theta=-\fr{1}{2Y^2}\left(\alpha((1-2\sigma)v-u)
                          +4\alpha\beta\chi+\beta((1+2\sigma)v-u))\right),\n
&\sigma_{ab}=-\fr{1}{6}(-2\de^{(uv)}_{ab}g_{uv}+
\de^{\theta\theta}_{ab}g_{\theta\theta}+\de^{\phi\phi}_{ab}g_{\phi\phi})\Theta,~~~
\sigma(n)=-\fr{\sqrt{10}}{24Y^2}\Theta,\n
&K_{ab}\equiv g^c_a n_{b;c}=-\fr{1}{2}
(\de^{\theta\theta}_{ab}g_{\theta\theta}+\de^{\phi\phi}_{ab}g_{\phi\phi})\Theta,
\label{tsk}
\ear
respectively.
The projected Weyl tensor is
\be
E_{ef}\equiv C_{acbd}n^cn^dg^a_eg^b_f,
\label{pweyl}
\ee
here
\be
E_{uu}=-\fr{1}{3}R_{uu},~~~
E_{vv}=-\fr{1}{3}R_{vv},~~~
E_{uv}=+\fr{1}{3}R_{uv}-\fr{1}{4}Rg_{uv},~~~
E_{\theta\theta}=g_{\theta\theta}(\fr{1}{4}R+\fr{2}{3}(1-\alpha\beta)R_{uv})
\ee

Transferring to single null coordinates using \ref{dncoord},
further changing coordinates using
\be
r'=r+\alpha\chi,~~~
v'=v+\fr{(\alpha-\beta)\chi}{2\sigma},~~~~~~~
\sigma\ne0,
\label{snab}
\ee
and dropping the primes,  the solution becomes
\ber
ds^2&=&-\fr{(1+2\sigma)}{(1-\alpha\beta)}dv^2+\fr{2}{(1-\alpha\beta)}dvdr
+r(r-2\sigma v)d\Sigma^2_2\n
&&+\fr{(\alpha-(1+2\sigma)\beta)}{\sigma(1-\alpha\beta)}dvd\chi
-\fr{(\alpha-\beta)}{\sigma(1-\alpha\beta)}drd\chi
+\left(1-\fr{(\alpha-\beta)\gamma}{4\sigma^2(1-\alpha\beta)}\right)d\chi^2,\n
\phi&=&\fr{1}{2}\ln\left(1-\fr{2\sigma v}{r}\right)
\label{es5d119}
\ear
with $\gamma$ given by \ref{gamma}.
Features of the line element in this form are that
1)it is of the same form as
the four dimensional case except for constant factors.
Truncating the line element \ref{es5d119}
by simply putting everything involving $\chi$ to zero,
gives non vanishing $R_{\theta\theta}=R_{\phi\phi}=\alpha\beta$
and simple relationships between the product invariants,
such as \ref{5dinvars}, are lost;
however taking one of $\alpha$ or $\beta$ to vanish gives back \ref{esnd4}.
Going back to the five dimensional case \ref{es5d119}
$\alpha=0,~\beta=1$ gives $\Theta=-r\exp(2\phi)/Y^2$
so that there is essentially one independent object the scalar field $\phi$,
the projected Weyl tensor \ref{pweyl} now does not have a dependence on $\chi$
but does not seem to be simply expressible in terms of the scalar field $\phi$,
2)$\chi$ does not appear explicitly in the metric,
3)the scalar field takes exactly the same form as in the four dimensional case,
4)there are no non-vanishing Riemann or Ricci tensor $\chi$ indexed components,
but there are Weyl tensor $\chi$ indexed components,  also no component depends on $\chi$.
The Ricci scalar is
\be
R=\left[(\alpha-\beta)^2r^2-4\sigma rv(2\sigma(1-\alpha\beta)+\alpha(\alpha-\beta))
+4\sigma^2v^2((1+2\sigma)(1-\alpha\beta)+\alpha^2)\right]/(2Y^4).
\ee
Using the same projection vector \ref{proj},
the acceleration and shear vanish,
the extrinsic curvature take the same form as in \ref{tsk}
the expansion is
\be
\Theta=\fr{2\alpha\sigma v-(\alpha+\beta)r}{Y^2}
\ee
and the shear is
\ber
&\sigma(n)=-\fr{\sqrt{10}}{12}\Theta,~~~
\sigma_{ab}=
-\fr{1}{6}(\de^{\theta\theta}_{ab}g_{\theta\theta}+\de^{\phi\phi}_{ab}g_{\phi\phi})\Theta\\
&+\fr{1}{3}
\left(\de^{rv}_{ab}g_{rv}+\de^{r\chi}_{ab}g_{r\chi}
+\de^{vv}_{ab}g_{vv}+\de^{v\chi}_{ab}g_{v\chi}
+\de^{\chi\chi}_{ab}g_{\chi\chi}
\fr{(\beta-\alpha)\gamma}{4\sigma^2(1-\alpha\beta)-(\alpha-\beta)\gamma}\right)\Theta,
\nonumber
\ear
the projected Weyl tensor \ref{pweyl} has $\chi$ components
and does not seem to be simply expressible.

For $\sigma=0$ the coordinate transformation \ref{snab} and the metric \ref{es5d119}
are not defined;
defining $t\equiv v-r$, \ref{e5d94} reduces to
\be
ds^2=
\fr{1}{1-\alpha\beta}(-dt^2+dr^2)+(r+\alpha\chi)(r+\beta\chi)d\Sigma^2_2
+d\chi^2,~~~
\phi=\fr{1}{2}\ln\left(\fr{r+\beta\chi}{r+\alpha\chi}\right),
\label{seo}
\ee
which does not seem to further simplify.
With respect to the vector field $t^a=\de^a_t$,  many Lie derivatives vanish,
in particular $\Li_tRie^2=0$.
For \ref{seo} when $\alpha=\beta$ the metric is flat.
\section{The second five dimensional solution.}
In double null coordinates the second five dimensional generalization
of the four dimensional solution is
\be
ds^2_5=-dudv+Y^2d\Sigma^2_2
+2\beta^2d\chi((1+2\sigma)dv+du)
+\gamma^2d\chi^2,
\label{e5d15}
\ee
with $Y^2$ given by \ref{ednd4} and
\be
\gamma^2\equiv\alpha^2(\chi)\gamma'((1-2\sigma)v-u)^2
\ee
$\gamma,\gamma'~\&~\alpha$ are functions,  $\gamma$ can be set to 1 or 0.
In the $\beta^2d\chi$ term the relative size of the du and dv contributions
is fixed by the requirement $R_{\theta\theta}=0$.
The scalar field is the same as for the four dimensional minimal scalar \ref{sfsn4},
the fifth component vanishing identically.
In general this does not seem to be related to the first five dimensional generalization
\ref{e5d94},
because of the factor of $1-\alpha\beta$ there.
The Ricci scalar is
\be
R=\fr{32\sigma^2((1+2\sigma)^2\beta^4v^2+(\alpha^2\gamma^2+2(1+2\sigma)\beta^4)+\beta^4u^2)}
{((1+2\sigma)v-u)^2((1-2\sigma)v-u)^2(\alpha^2\gamma^2+4(1+2\sigma)\beta^4)}
\ee
The product invariants are again completely determined by the Ricci scalar,
and are given by \ref{5dinvars}.
Projecting using \ref{proj} the acceleration and rotation vanish
and the expansion,  extrinsic curvature and shear are
\ber
&\Theta=\fr{8\sigma\beta^2}{(u-(1-2\sigma)v)(\alpha^2\gamma^2+4(1+2\sigma)\beta^4)},~~~
K_{\theta\theta}=\sin(\theta)^{-2}K_{\phi\phi}=
\fr{u-(1+2\sigma)v}{8(u-(1-2\sigma)v)}\Theta,\\
&\sigma_{ab}=
-\fr{1}{3}\left(\fr{\alpha^4\gamma^4+4(1+2\sigma)\alpha^2\gamma^2\beta^4-1}
                   {\alpha^2\gamma^2+4(1+2\sigma)\beta^4}\de_{ab}^{\chi\chi}
+2\de_{ab}^{(uv)}g_{uv}+\de_{ab}^{u\chi}g_{u\chi}+\de_{ab}^{v\chi}g_{v\chi}
-\fr{1}{8}
(\de_{ab}^{\theta\theta}g_{\theta\theta}+\de_{ab}^{\phi\phi}g_{\phi\phi})\right)\Theta
\nonumber
\ear
all of which vanish for $\beta=0$.   The shear scalar and
the projected Weyl tensor are independent of $\chi$ and do not seem to simply factor.
\section{The third five dimensional solution.}
The solution is
\be
ds^2_5=\sqrt{\chi}\left\{-dudv+Y^2d\Sigma^2_2\right\}+d\chi^2,~~~
\phi_2=\fr{1}{2}\sqrt{\fr{3}{2}}\ln\chi,
\label{e5d26}
\ee
with the term in the brackets given by \ref{esnd4}
and $\phi_1$ given by \ref{dn4s}.
The conformal factor $\sqrt{\chi}$ fixed by requirement $R_{\theta\theta}=0$.
This is a solution for two scalar fields to the field equations
\be
R_{ab}=2\phi_{1,a}\phi_{1,b}+2\phi_{2,a}\phi_{2,b}.
\label{twosf}
\ee
The N\"other current is
\be
j_a\equiv i(\phi^*\p_a\phi-\phi\p_a\phi^*)
=\phi_2\phi_{1a}-\phi_1\phi_{2a}.
\ee
Here the N\"other current is
\be
j_a=\fr{\sqrt{6}}{8}\left[\fr{\sigma\sqrt{\chi}\ln(\chi)}{Y^2}(-v\de^u_a+u\de^v_a)
-\fr{1}{\chi}\ln\left(\fr{(1-2\sigma)v-u}{(1+2\sigma)v-u}\right)\right].
\label{noec}
\ee
The size of N\"other current is
\be
j_a^2=-\fr{1}{2\sqrt{\chi}}R^{(4)}\phi_2^2+\fr{3}{8\chi^2}\phi_1^2,
\label{nesz}
\ee
and this is of undetermined sign,
so that the N\"other current can be timelike,  null,  or spacelike.

The scalar invariants are
\ber
&R=\fr{R^{(4)}}{\sqrt{\chi}}+\fr{3}{4\chi^2},~~~
Weyl^2=\fr{11R^{(4)^2}}{6\chi},\\
&Rie^2=\fr{3R^{(4)^2}}{\chi}-\fr{R^{(4)}}{4}\chi^{-\fr{5}{2}}+\fr{21}{32\chi^4},~~~
Ricci^2=\fr{R^{(4)^2}}{\chi}+\fr{9}{16\chi^4}.\nonumber
\ear
where $R^{(4)}$ is given by \ref{rs4}.

Using the projection vector \ref{proj},
the acceleration and rotation vanish
and the expansion,  extrinsic curvature,  shear,  and projected Weyl tensor,
which for this metric is the same as the electric part of the Weyl tensor are
\ber
&\Theta=\fr{1}{\chi},~~~
\sigma(n)=\fr{1}{6\chi}\\
&K_{ab}=\fr{1}{4}(\de_{ab}^{uv}g_{uv}
+\de_{ab}^{\theta\theta}g_{\theta\theta}+\de_{ab}^{\phi\phi}g_{\phi\phi}),~~~
\sigma_{ab}=-\fr{1}{3}K_{ab},\n
&E_{ab}=\fr{\sigma^2\chi}{6Y^4}(-v^2\de_{ab}^{uu}+uv\de_{ab}^{(uv)}-u^2\de_{ab}^{vv})
+\fr{\sigma^2uv\sqrt{\chi}}{6}
(\de_{ab}^{\theta\theta}g_{\theta\theta}+\de_{ab}^{\phi\phi}g_{\phi\phi})\nonumber
\label{noexv}
\ear
\section{Five dimensional Vaidya spacetime.}
A generalization of Vaidya's spacetime to five dimensions is
\be
ds^2=\sqrt{\chi}\{-\left(1-\fr{2m(v,\chi)}{r}\right)dv^2+2dvdr+r^2d\Sigma^2_2\}+d\chi^2,
\label{vaid5}
\ee
choosing the requirement that $R_{\theta\theta}=0$
fixes the conformal factor as $\sqrt{\chi}$,
\ref{vaid5} has Ricci tensor
\be
R_{vv}=\fr{2m_v}{r^2}-\fr{(\chi m_\chi)_\chi}{r\sqrt{\chi}},~~~
R_{rv}=\fr{m_\chi}{r^2},~~~
R_{\chi\chi}=\fr{3}{4\chi^2},
\ee
the $\chi$ dependence of $m$ means that the four dimensional stress is no longer that
of a null radiation field.  The $R_{\chi\chi}$ component is non vanishing even when
$m$ is independent of $\chi$.
The Ricci scalar and product invariants are
\be
R=\fr{3}{4\chi^2},~~~
Weyl^2=\fr{48m^2}{r\chi},~~~
Ricci^2=R^2,~~~
Rie^2=Weyl^2+\fr{7}{6}R^2,
\label{vadinv}
\ee
these do not explicitly involve derivatives of m with respect to either $v$ or $\chi$.
Using the projection vector \ref{proj},
the acceleration and rotation vanish
and the expansion,  extrinsic curvature,  shear,  and projected Weyl tensor,
which for this metric is the same as the electric part of the Weyl tensor are
\ber
&\Theta=\fr{1}{\chi},~~~
K_{vr}=\fr{1}{4\sqrt{\chi}},~~~
K_{\theta\theta}=\fr{r^2}{4\sqrt{\chi}},\\
&\sigma=\fr{1}{6\chi},~~~
\sigma_{rv}=-\fr{1}{12\sqrt{\chi}},~~~
\sigma_{\theta\theta}=-\fr{r^2}{12\sqrt{\chi}},~~~
\sigma_{vv}=\fr{1}{12r\sqrt{\chi}}(12\chi m_\chi+r-2m),\n
&E_{vv}=-\fr{2\chi^{\fr{1}{4}}}{3r}(\chi^{\fr{1}{4}}m_\chi)_\chi-\fr{2}{3r^2}m_v.\nonumber
\ear
\section{Conclusion.}
There at least three criteria one could use to test interaction with the fifth dimension.
The {\it first} is {\bf explicit interaction}.
The Einstein tensor for a minimal scalar field is $G_{ab}=2\phi_a\phi_b-\phi_c^2g_{ab}$,
so that with respect to a vector field $v_a$ there is the momentum transfer
$\pi_a=v^bG_{ab}=2\phi_a\phi\cdot v-v_a\phi_c^2$.
Such a momentum transfer seems unavoidable for non vanishing scalar field because of
the metric (or second) term in the Einstein tensor;
whether this is good or bad depends on ones point of view.
It is good if one simply wants any indication of transfer of information.
It is bad if one wants only gravity,  in the sense that $R_{5a}=0$,
to be present in the fifth dimension,
because there will also be the scalar field present.
For the above examples it is also bad because there is no neat way of characterizing
the energy transfer,  it is not even clear when it will be timelike,  null, or spacelike.
The {\it second} is {\bf implicit interaction}.
By this is meant that the four dimensional metric takes a different form than would
be expected from four dimensional theory and that this difference can somehow be measured.
To illustrate this consider \ref{es5d119},  except for the factor of $1-\alpha\beta$ the
metric truncated to four dimensions would be of the same form as \ref{esnd4}.
Roughly the $dv^2$ term suggests a change in the null velocity from
$c\rightarrow c/\sqrt(1-\alpha\beta)$;
however other metric terms change as well and it turns out not to be possible to have
only the null velocity change occurring.
For $\sigma=0$, given by equation \ref{seo}, similar problems apply.
The {\it third} is {\bf N\"other criteria}.
One could imaging scalar fields as in someway corresponding to a quantum mechanical
wave function of some part of a system.
A quantum mechanical interaction might be indicated by a non vanishing N\"other current
between one part of the system and another.
To model this one would need exact solutions for two or more scalar fields.
It turns out to be simple to produce solutions for linear combinations of scalar fields,
however these have vanishing N\"other current.
Spherical symmetry imposes a high degree of symmetry making finding solutions with
a non vanishing N\"other current hard to find.
An example with a non vanishing N\"other current is \ref{e5d26};
but properties of this solution include
1) the two scalar fields are disconnected,  in the sense that one scalar field
depends on one set of coordinates $\{r,v\}$ and the other depends on $\{\chi\}$,
2) the current can be timelike,  null or spacelike,
and
3)there is no simple way of characterizing what happens to any N\"other charge.
Another problem in general with the N\"other criteria is that in quantum cosmology the high
degree of symmetry means that there are no N\"other currents.
To conclude the five dimensional scalar-Einstein equations provide simple exact
solutions with which to discuss interaction with the fifth dimension,
however their interpretation is difficult.

\end{document}